\documentclass[pra,twocolumn,superscriptaddress,amssymb,floatfix,10pt,aps]{revtex4-1}
\usepackage{graphicx}
\usepackage{amsmath,amsfonts,amssymb,bm}
\usepackage[english,spanish]{babel}
\usepackage{url}

\begin{document}

\title{Enhanced light-harvesting of protein-pigment complexes assisted by a quantum dot antenna}

\author{Gabriel Gil}
\affiliation{Dipartimento di Scienze Chimiche, Universit\`a degli studi di Padova, Italy \\
Instituto de Cibern{\'e}tica, Matem{\'a}tica y F{\'i}sica (ICIMAF), Havana, Cuba\\
gabrieljose.gilperez@unipd.it}
\author{Guido Goldoni}
\affiliation{Dipartimento di Scienze Fisiche, Informatiche e Matematiche, Universit{\`a} degli studi di Modena e Reggio Emilia, Modena, Italy \\
CNR Istituto di Nanoscienze, Centro S3, Modena, Italy\\
guido.goldoni@unimore.it}
\author{Stefano Corni}
\affiliation{Dipartimento di Scienze Chimiche, Universit{\`a} degli studi di Padova, Italy \\
CNR Istituto di Nanoscienze, Centro S3, Modena, Italy\\
stefano.corni@unipd.it}

\begin{abstract}
We predict the enhanced light harvesting of a protein-pigment complex when assembled to a quantum dot (QD) antenna. Our prototypical nanoassembly setup is composed of a Fenna-Mattews-Olson system hosting 8 Bacteriochlorophyll (BChl) a dyes, and a near-infrared emitting CdSe$_x$Te$_{(1-x)}$/ZnS alloy-core/shell nanocrystal. BChl a has two wide windows of poor absorption in the green and orange-red bands, precisely where most of the sunlight energy lies. The selected QD is able to collect sunlight efficiently in a broader band and funnel its energy by a (non-radiative) F\"orster resonance energy transfer mechanism to the dyes embedded in the protein. By virtue of the coupling between the QD and the dyes, the nanoassembly absorption is dramatically improved in the poor absorption window of the BChl a. 
\end{abstract}

\maketitle

\section{Introduction}

Photosynthesis is a unique process by which most plants and algae, as well as cyanobacteria, convert solar light into chemical energy \cite{collini-etal,linnanto-korppi}. In the first two steps of a photosynthetic pathway, solar energy is collected and funneled into reaction centers, where it triggers a series of chemical transformations leading to the storage of energy in the form of chemical compounds, such as ATP. The photochemical agents responsible for those two basic steps are pigment molecules hosted by proteins resident in cells or biological membranes \cite{collini-etal,linnanto-korppi,milder-etal}. Pigments are excited by light and they assume the donor or acceptor role in a chain of F\"orster resonance energy-transfer (FRET) processes that ultimately delivers the excitation energy into other pigments, within the reaction center protein, able to produce an electron transfer \cite{collini-etal,linnanto-korppi}. 

Despite its many intricacies, photosynthesis is ranked among the most efficient of natural mechanisms \cite{janssen-etal}. Almost every photon absorbed by the pigments leads to an electron transfer at the end of the energy-transfer chain (about 100\% of quantum efficiency) \cite{mirkovic-etal}, and the overall energy conversion yield is near 34\% under ideal conditions \cite{rabinowitch-govindjee}. However, there is room for improvement. Pigments are only able to capture a small fraction of sunlight photons due to their narrow absorption bands and their far from ideal overlap with the solar emission spectrum. As an instance, the excitation spectrum of BChl a --a light-harvesting specialized pigment found in photosynthetic bacteria-- features two wide windows of poor absorption between 400-800 nm, coincident with highest contributions to the solar emission spectrum. Therefore, a promising way of boosting the performance of photosynthesis is by enhancing the efficiency of the very first light-harvesting stage. This idea encourages many artificial photosynthesis applications. For example, recently, it was proven experimentally that at least a threefold increase in the electron transfer rate can be achieved by coupling a quantum dot (QD) antenna directly to a reaction center protein-pigment complex (from \textit{Rodobacter Sphaeroides}) devoided from its light-harvesting dyes \cite{nabiev-etal}. In the same spirit, although outside photosynthesis context, the same group have demonstrated an improvement in the light-driven biological operation of a retinal chromophore in a Bacteriorhodopsin protein (from \textit{Halobacterium salinarum}) when integrated with a QD antenna \cite{rakovich-etal}.

QDs are inorganic semiconductor nanostructures characterized by a wide absorption spectrum (of ever increasing absorbance toward shorter wavelength) that can be tuned with the size and chemical composition to obtain optimal overlap with the solar emission spectrum \cite{alivisatos}. These features render QDs excellent candidates for enhanced light-harvesting applications. Furthermore, when suitably coupled with protein-pigment complexes, QDs can also transfer their excitation energy efficiently (through FRET) to the pigments embedded in the protein scaffold.

The abovementioned rationale can explain qualitatively the remarkable achievement presented in Refs.~\cite{nabiev-etal}. However, a more fine-grained theoretical modeling/simulation of such nanoassembly can shed light onto its operation and moreover it can suggest how to engineer even better performances. For instance, the specific QD employed in Ref.~\cite{nabiev-etal}, however globally spherical, has an intrinsic anisotropic lattice (wurtzite crystal system) which must influence dramatically the FRET rate to the pigments, depending on the relative location and orientation of subsystems involved (i.e., the protein-pigment complex and the QD) \cite{ourJCP,ourRSCAdv}. Therefore, by placing adequately the protein-pigment complex with respect to the QD one might tune its overall QD-protein energy-transfer efficiency to its highest value. We have recently shown \cite{ourJCP,ourRSCAdv} that prediction of optimal quantum efficiency can be readily obtained by a theoretical model which takes into account the position and orientation of the fragments in the nanoassembly, the atomistic structure of the dyes and the crystal symmetries QDs in the transfer rates entering a kinetic equation approach. Our approach also provides a clear physical picture of the results.

On the other hand, Ref.~\cite{nabiev-etal} uses a nanoassembly in which the reaction center protein lacks the pigments in charge of solar energy collection. The energy-transfer photosynthetic pathway in photoautotroph organisms is a nearly flawless product of natural selection, where the relative orientation and positions of dyes and their local environment within proteins, etc., appear to be optimized to give rise to very high quantum efficiencies (i.e., one electron transfer in output per each excitation input). Specialized light-harvesting pigments might not only be important for their photoabsorption capabilities but they might also take part in driving the excitation toward the reaction center. It is hence desirable to keep intact such a fine-tuned photosynthetic pathway while enhancing its light-harvesting deficiencies. For instance, we may envisage an application in which the QD antenna funnels its energy directly toward the light-harvesting pigments and not to other pigments located subsequently within the energy-transfer chain, leaving the natural photochemical path intact.

Inspired by the work of Nabiev et al.,\cite{nabiev-etal} in this contribution our goal is twofold: (i) to present a theoretical approach realistic enough to capture the relevant phenomenology of light harvesting and excitation energy transferring within the nanoassembly, and (ii) to show that a dramatic enhancement of the solar energy collection is feasible by assembling a QD antenna together with protein-pigment complex without need to suppress of the specialized light-harvesting pigments from their encapsulating protein.

\section{Materials and methods}

\begin{figure}[ht]
\centering
\includegraphics[width=\linewidth]{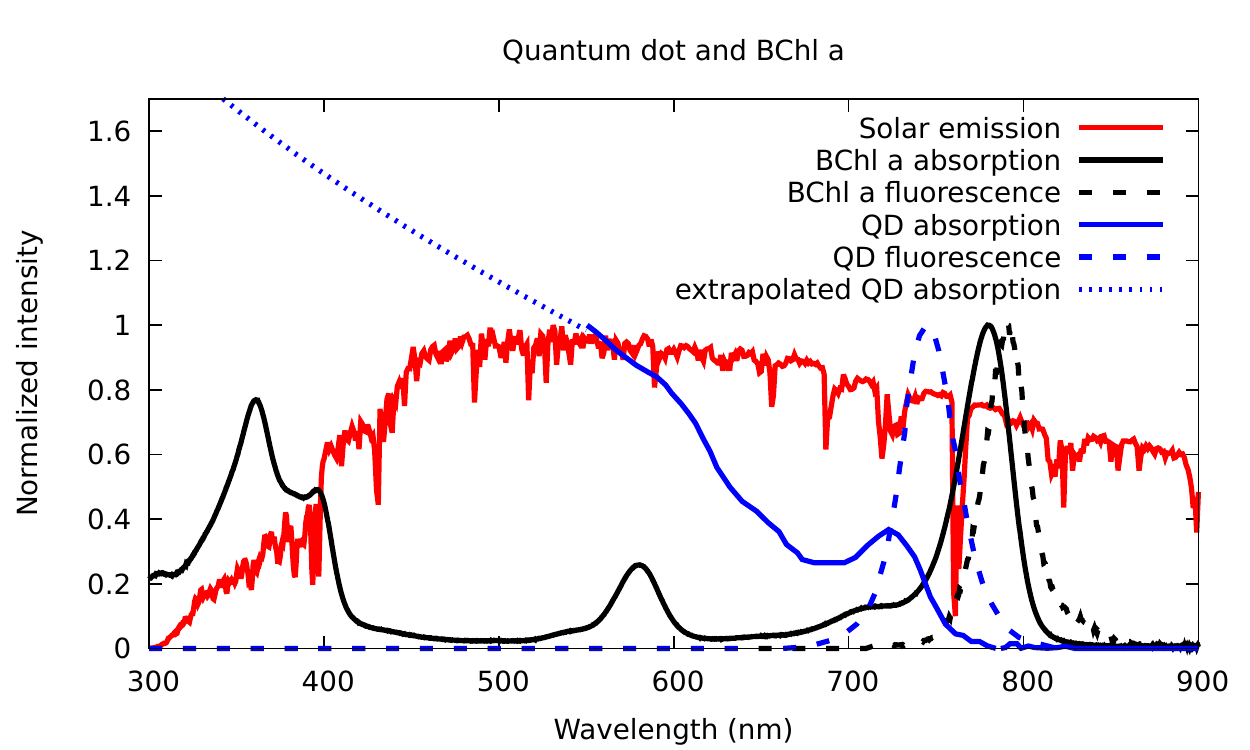}
\caption{Experimental absorption and emission spectra of BChl a \cite{bchla-data,connelly-etal} and the QD-WS-yy \cite{qdot-data}, together with the solar emission spectrum \cite{sunlight-data}. The extrapolation of QD absorption spectrum based on a direct gap bulk semiconductor analytical formula \cite{rosencher-vinter} (see text) is overlaid for wavelength $\le 550$ nm.}
\label{fig_overlap}
\end{figure}

To that aim, we choose a prototypical system composed of a Fenna-Mattews-Olson (FMO) protein accomodating 8 BChl a dyes \cite{milder-etal,tronrud-etal}, and a near-infrared emitting CdSe$_x$Te$_(1-x)$/ZnS alloy-core/shell semiconductor nanocrystal (also known as quantum dot) \cite{qdot-data}. The specific FMO protein comes from \textit{Prosthecochloris Aestuarii} bacterial organism and its structure (taken from the Protein Data Bank) is known from X-ray diffraction with 1.3 \AA~ resolution \cite{tronrud-etal,pdb,fmo-data}. From this protein structure, we extract the geometry of the BChl a dyes and their relative locations within the FMO volume. Other relevant data, such as the absorption and emission spectra, as well as the quantum yield, of isolated BChl a in toluene solution are taken from PhotochemCAD package (see Fig.~\ref{fig_overlap}) \cite{du-etal,dixon-etal,connelly-etal}. Ref.~\cite{milder-etal} concluded that only one or two of the BChl a molecules contribute significantly to excited states wavefunction in FMO. In the present work, we approximate the excited states as the excited states of the uncoupled BChl a. We perform Linear Response Time-Dependent Density Functional Theory (LR-TDDFT) calculations on one of BChl a from the FMO in vacuo using the General Atomic and Molecular Structure System (GAMESS) \cite{GAMESS}, and we extract the transition dipole moment corresponding to the excited state associated with the first peak in the experimental absorption spectrum. We model the FMO as a collection of BChl a dyes, disregarding any protein-pigment interactions, however maintaining the spatial position and orientation fixed by the protein scaffold. Furthermore, each BChl a molecule (e.g., the $i$-th) is represented by its transition dipole moment ($\mathbf{d}_{\mathrm{M}_i}$) with different orientation (but the same magnitude) sitting on a characteristic position ($\mathbf{R}_{\mathrm{M}_i}$) that we estimate from the central Mg atom coordinates. This transition dipole moment enters in the calculation of the inter-pigment and the QD $\rightarrow$ pigment (or viceversa) electronic coupling, in turn an essential ingredient in FRET rate (see Ref.~\cite{ourJCP,ourRSCAdv} for details).

The specific QD has a core/core+shell radius of 1.95/2.85 nm, it emits at 740 nm wavelength and it is commercially available from Nomcorp company with the code name of QD740-WS-yy \cite{qdot-data}. We take the absorption and emission spectra of the latter QD in water from the producer webpage (see Fig.~\ref{fig_overlap}) \cite{qdot-data}. Experimental absorption is sampled roughly down to 550 nm; for lower wavelength, we extrapolate the absorption spectrum on the grounds of an analytical formula valid for direct gap bulk semiconductors \cite{rosencher-vinter}, fitted to the 5 data points closer to 550 nm (see Fig.~\ref{fig_overlap}). Data on the specific stoichiometry of the alloy core (i.e., CdSe$_{0.34}$Te$_{0.66}$) have been taken from Ref.~\cite{bailey-nie}. Bulk structures of CdSe, CdTe and ZnS suggest that the underlying QD lattice is cubic (zincblende) \cite{madelung}. This symmetry generates a threefold degeneracy in the interband transition dipole moment of the QD, reflecting on favoured QD $\rightarrow$ dye FRET when the dye transition dipole moment orientation and position are aligned with specific QD crystal axes \cite{bastard}. Anyhow, being more symmetric, zincblende should exhibit less anisotropy in the QD $\rightarrow$ dye FRET than wurtzite (hexagonal crystal system) \cite{ourJCP,ourRSCAdv}. We estimate the effective masses of quasi-particle electrons and heavy holes and dielectric constant of the alloy core, together with the valence and conduction band offset between core and shell materials, from a stoichiometry-based linear interpolation of analogous data for CdSe and CdTe bulk materials \cite{madelung,hinuma-etal}. The fluorescence wavelength of typical-size CdSe and CdTe QDs reach up to $\sim$670 nm \cite{bailey-nie}. Alloyed CdSe$_{x}$Te$_{(1-x)}$ QD of similar size can push further the emission wavelength to the near infrared (NIR) thanks to the optical bowing effect \cite{bailey-nie,wei-etal}. Using NIR emitting QDs is essential for the prototypical nanoassembly we are focusing on. It comes as constraint from the protein-pigment complex side to maximize the spectral overlap ($J_{\mathrm{QD}\rightarrow\mathrm{M}}$) between the BChl a first absorption band at nearly 800 nm (see Fig.~\ref{fig_overlap}) and the fluorescence peak from the QD. The FRET rate is proportional to the spectral overlap between donor and acceptor of energy involved in the transferring event. Better spectral overlaps of such kinds leads to larger QD $\rightarrow$ pigment FRET rates and therefore better harvesting of light by the FMO. We model the QD within an effecttive mass approach assuming a spherical shape, finite and infinite square confinement potentials for core and shell materials, respectively, and a water environment. We perform Configuration Interaction calculations for an electron-hole pair to calculate the excitonic states of the QD (see Ref.~\cite{ourJCP}). The position-dependent transition electric field ($\mathbf{E}_{\mathrm{QD}}(\mathbf{r})$) associated with the first exciton peak in the experimental absorption spectrum can be directly computed from the first excited state of the electron-hole pair. Finally, the rate of the FRET from QD to the $i$-th pigment is calculated as $k_{\mathrm{QD}\rightarrow\mathrm{M}_i}(\mathbf{r})=2\pi J_{\mathrm{QD}\rightarrow\mathrm{M}} |\mathbf{d}_{\mathrm{M}_i}\cdot\mathbf{E}_{\mathrm{QD}}(\mathbf{R}_{\mathrm{M}_i})|^2$.

\begin{figure}[ht]
\centering
\includegraphics[width=0.5\linewidth]{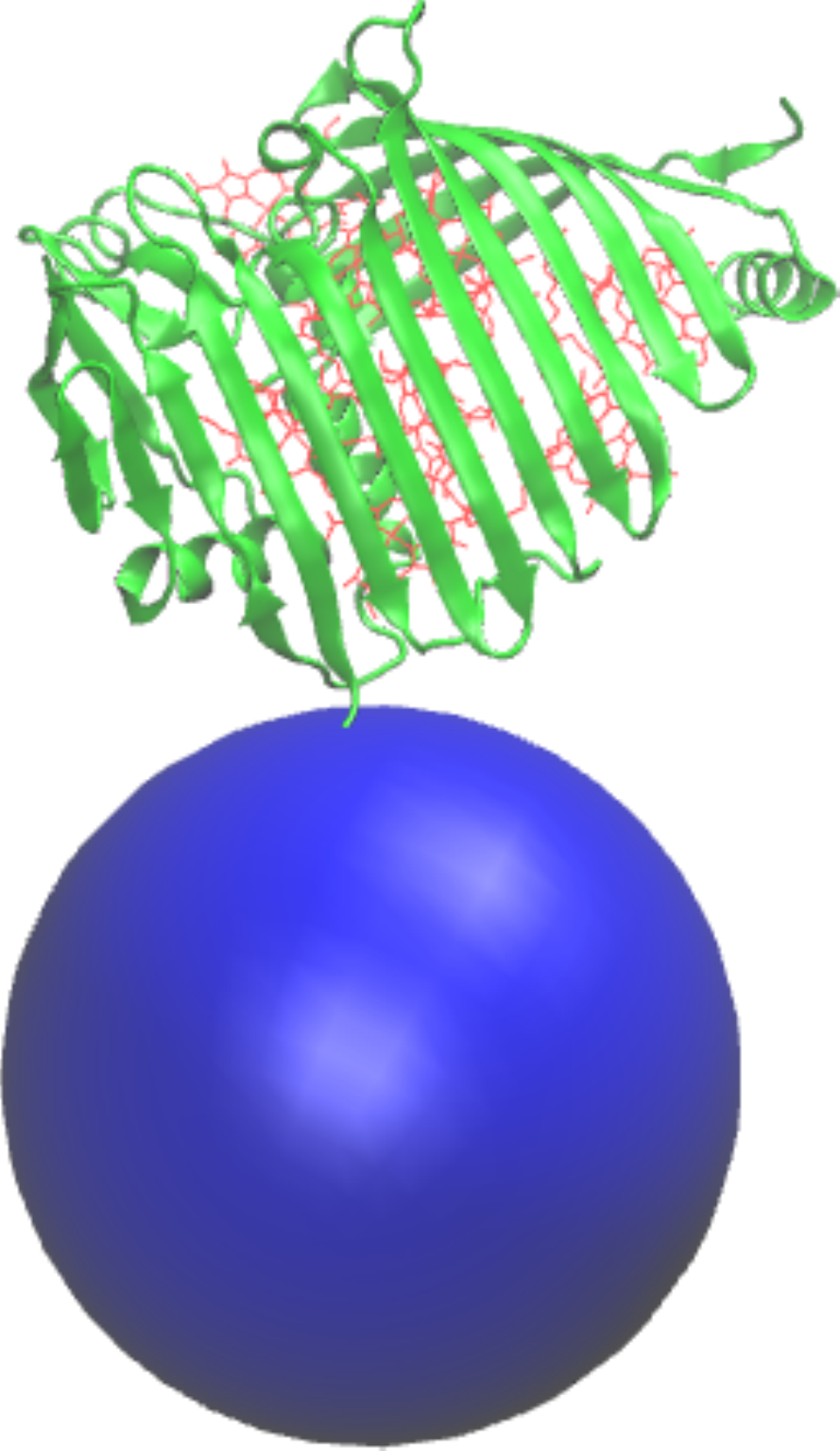}
\caption{Schematics of a nanoassembly setup composed of a FMO protein (backbone in green) encapsulating 8 BChl a dyes (molecular structure in red) and a quantum dot (drawn as a blue sphere). The quantum dot size (diameter 5.7 nm) is represented to scale.}
\label{fig_FMONP}
\end{figure}

Our theoretical description relies on a kinetic model for the evolution of the excited state population in the pigments and the QD, considering competitive absorption, FRET and decay channels (as in Ref.~\cite{ourRSCAdv}). All QD - pigment, pigment - QD, as well as inter-pigment energy transfer are included. The absorption rates consider explicitly the wavelength dependence of the QD and the dyes spectra (see Fig.~\ref{fig_overlap}). A detailed account on kinetic model and the calculation of rates for each of these processes have been presented elsewhere (see Ref.~\cite{ourJCP,ourRSCAdv}). We calculate the photoexcitation spectrum of the dyes from the total fluorescence rate at the steady state as a function of the absorption wavelength. The photoexcitation spectrum quantify how much fluorescence intensity is obtained at any excitation wavelength, and when the analyzed system is isolated (i.e., not coupled to any other fragment) it is proportional to the absorption spectrum. When a fragment is coupled to some other that can act as an extra source for the excitation (apart from light itself), the photoexcitation spectrum measure the overall capacity of the coupled system to absorb light and transfer its energy to the fragment under examination. The QD and FMO are placed at a contact distance, with threonine 1 residue practically on top of the QD surface and the closer BChl a about 1.4 nm away from it (see Fig.~\ref{fig_FMONP}).

\section{Results}

\begin{figure}[ht]
\centering
\includegraphics[width=\linewidth]{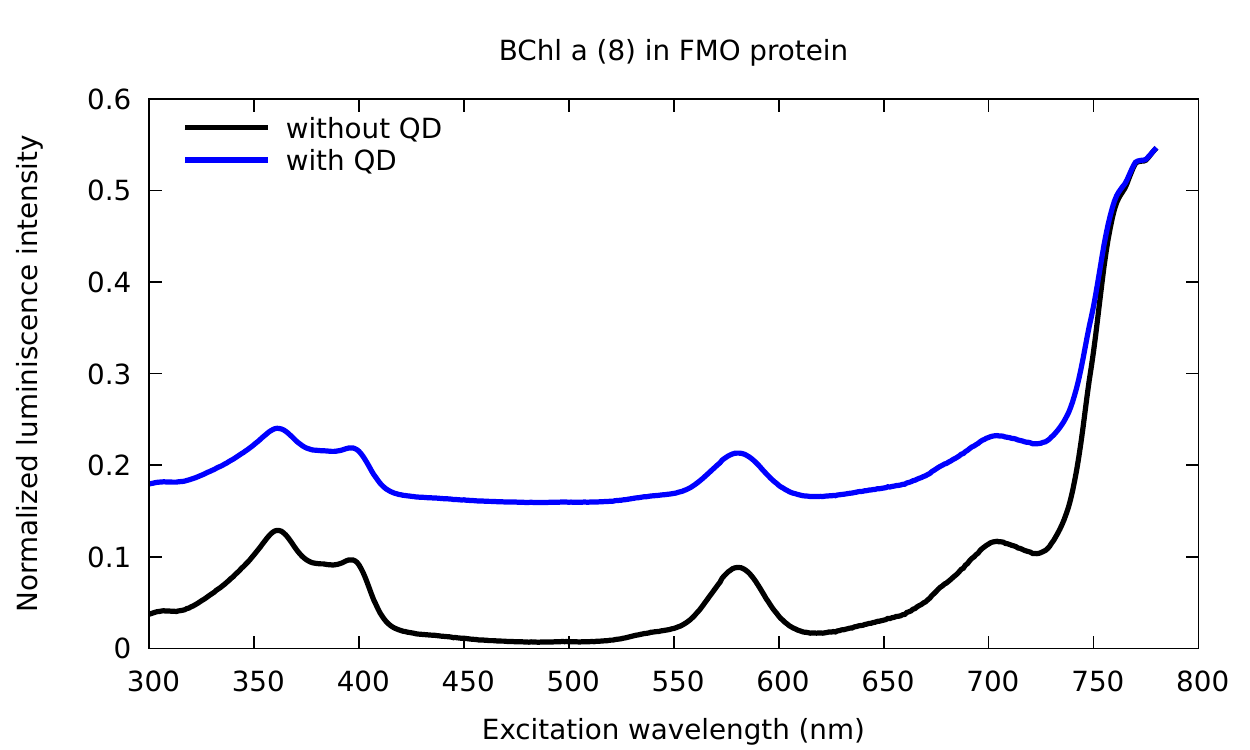}
\caption{Photoexcitation spectra of BChl a within a FMO complex, isolated (black line) or in presence of a QD antenna (blue line).}
\label{fig_spectra}
\end{figure}

Fig.~\ref{fig_spectra} shows our result for the photoexcitation spectra of FMO set of BChl a isolated or in presence of the QD. A careful observation of figure show that the absorption peaks of the single isolated dyes are reproduced in both of the photoexcitation spectra (with and without the QD antenna) although with different relative intensities. However, the most important result in this plot is the fact that deficient absorption energy regions feature a tremendous gain in absorbance when the QD antenna is considered. In fact, we compute a $\sim$2.6 times increase in the full photoexcitation intensity (integrated over a span of 300-780 nm) when the QD is coupled to the FMO. This enhancement in absorption is not a scaling factor and it depends on the wavelength. For instance, no increase in absorbance is obtain at nearly 800 nm, where the BChl a is the only active light-harvester, whereas nearly 22 times increase is obtained at 500 nm, where solely the QD absorb (cf. Fig.~\ref{fig_overlap}).

\section{Conclusions}

In summary, we predict that a NIR QD assembled to a FMO protein-pigment complex should overcome the hampered photoabsorption of the protein-embedded BChl a dyes in wide bands of the visible spectrum. In particular, one can expect an over 20 times increase in the fluorescence from the BChl a for green excitation photons. The nearby QD acts as (i) an efficient and complementary light collector, thanks to its wide absorption band, and (ii) as an additional source for the excitation energy, delivered to the FMO through efficient non-radiative energy transfer (FRET). The theoretical model supporting such prediction --simple but physically sound-- is based on a rate equation incorporating the key mechanisms of FRET from the QD to the FMO, as well as other competitive excitation and de-excitation pathways. Our modeling can be promptly used to find the relative position and orientation of the nanoassembly components associated with optimal QD - FMO FRET, and therefore, might help establishing design principles for a device with even better light-harvesting performances.

\section{Acknowledgments}\nonumber

We acknowledge financial support from European Research Council (ERC) under the European Union's Horizon 2020 research and innovation programme, through the project TAME-Plasmons (Grant agreement No. 681285). We are grateful to our former co-workers A. Bertoni, for kindly offering his $\mathcal{CI}$tool code to perform Configuration Interaction calculations for the quantum dot, and A. Delgado, for useful discussions on the subject.

\bibliographystyle{unsrt}
\bibliography{mybib}

\begin{thebibliography}{10}

\bibitem{collini-etal}
E.~Collini, C.~Curutchet, T.~Mirkovic, and {G.D.} Scholes.
\newblock Electronic energy transfer in photosynthetic antenna systems.
\newblock In I.~Burghardt et~al. (eds.), editor, {\em Energy Transfer Dynamics
  in Biomaterial Systems}, pages 3--34. Springer-Verlag, Berlin Heidelberg,
  2009.

\bibitem{linnanto-korppi}
J.~Linnanto and J.~Korppi-Tommola.
\newblock Quantum chemical simulation of excited states of chlorophylls,
  bacteriochlorophylls and their complexes.
\newblock {\em Phys. Chem. Chem. Phys.}, 8:663--687, 2006.

\bibitem{milder-etal}
{M.T.W.} Milder, B.~Br{\"u}ggemann, R.~van Grondelle, and {J.L.} Herek.
\newblock Revisiting the optical properties of the fmo protein.
\newblock {\em Photosynth. Res.}, 104:257--274, 2010.

\bibitem{janssen-etal}
{P.J.D.} Janssen, {M.D.} Lambreva, N.~Plumer{\'e}, C.~Bartolucci, A.~Antonacci,
  K.~Buonasera, {R.N.} Frese, V.~Scognamiglio, and G.~Rea.
\newblock Photosynthesis at the forefront of a sustainable life.
\newblock {\em Frontiers in Chemistry}, 2(36), 2014.

\bibitem{mirkovic-etal}
T.~Mirkovic, {E.E.} Ostroumov, {J.M.} Anna, R.~van Grondelle, Govindjee, and
  {G.D.} Scholes.
\newblock Light absorption and energy transfer in the antenna complexes of
  photosynthetic organisms.
\newblock {\em Chem. Rev.}, Article ASAP, 2016.

\bibitem{rabinowitch-govindjee}
E.~Rabinowitch and Govindjee.
\newblock {\em Photosynthesis}.
\newblock Wiley, 1969.

\bibitem{nabiev-etal}
I.~Nabiev, A.~Rakovich, A.~Sukhanova, E.~Lukashev, V.~Zagidullin, V.~Pachenko,
  {Y.P.} Rakovich, {J.F.} Donegan, {A.B.} Rubin, and {A.O.} Govorov.
\newblock Fluorescent quantum dots as artificial antennas for enhanced light
  harvesting and energy transfer to photosynthetic reaction centers.
\newblock {\em Angew. Chem. Int. Ed.}, 49:7217--7221, 2010.

\bibitem{rakovich-etal}
A.~Rakovich, Bouchonville~N. Sukhanova, A., E.~Lukashev, V.~Oleinikov,
  M.~Artemyev, V.~Lesnyak, N.~Gaponik, M.~Molinari, M.~Troyon, {Y.P.} Rakovich,
  {J.F.} Donegan, and I.~Nabiev.
\newblock Resonance energy transfer improves the biological function of
  bacteriorhodopsin within a hybrid material built from purple membranes and
  semiconductor quantum dots.
\newblock {\em Nano Lett.}, 10:2640--2648, 2010.

\bibitem{alivisatos}
{A.P.} Alivisatos.
\newblock Perspectives on the physical chemistry of semiconductor nanocrystals.
\newblock {\em J. Phys. Chem.}, 100:13226--13239, 1996.

\bibitem{ourJCP}
G.~Gil, S.~Corni, A.~Delgado, A.~Bertoni, and G.~Goldoni.
\newblock Excitation energy-transfer in functionalized nanoparticles: Going
  beyond the f{\"o}rster approach.
\newblock {\em J. Chem. Phys.}, 144(7):074101, 2016.
\newblock Erratum: \textit{J. Chem. Phys.}, 2016, \textbf{145}, 089902.

\bibitem{ourRSCAdv}
G.~Gil, S.~Corni, A.~Delgado, A.~Bertoni, and G.~Goldoni.
\newblock Predicting signatures of anisotropic resonance energy transfer in
  dye-functionalized nanoparticles.
\newblock {\em RSC Adv.}, 2016.

\bibitem{tronrud-etal}
Wen J. Gay L. Blankenship~{R.E.} Tronrud, {D.E.}
\newblock The structural basis for the difference in absorbance spectra for the
  fmo antenna protein from various green sulfur bacteria.
\newblock {\em Photosynth. Res.}, 100:79--87, 2009.

\bibitem{qdot-data}
Qd740-ws-yy, near infrared water soluble cdsete/zns core shell quantum dots,
  740 nm.
\newblock \url{http://nomcorp.com/qd740-ws-yy-specifications-spectra/}.
\newblock Nomcorp, Accessed: 2018-10-09.

\bibitem{pdb}
Z.~Feng G. Gilliland T.N. Bhat H. Weissig I.N. Shindyalov P.E.~Bourne
  H.M.~Berman, J.~Westbrook.
\newblock The protein data bank.
\newblock {\em Nucleic Acids Research}, 28:235--242, 2000.

\bibitem{fmo-data}
Fenna-mattews-olson protein-pigment complex from prosthecochloris aestuarii 2k
  at 1.3~\aa~ resolution.
\newblock \url{https://www.rcsb.org/structure/3eoj}.
\newblock PDB ID: 3EOJ, Accessed: 2018-10-09.

\bibitem{du-etal}
R.-C. A. Fuh J. Li L. A.~Corkan Du, H. and J.~S. Lindsey.
\newblock Photochemcad: A computer-aided design and research tool in
  photochemistry.
\newblock {\em Photochem. Photobiol.}, 68:141--142, 1998.

\bibitem{dixon-etal}
M.~Taniguchi Dixon, J.~M. and J.~S. Lindsey.
\newblock Photochemcad 2. a refined program with accompanying spectral
  databases for photochemical calculations.
\newblock {\em Photochem. Photobiol.}, 81:212--213, 2005.

\bibitem{connelly-etal}
E.~B.~Samuel Connolly, J.~S. and A.~F. Janzen.
\newblock Effects of solvent on the fluorescence properties of
  bacteriochlorophyll a.
\newblock {\em Photochem. Photobiol.}, 36:565--574, 1982.

\bibitem{GAMESS}
M.W. Schmidt, K.K. Baldridge, J.A. Boatz, S.T. Elbert, M.S. Gordon, J.H.
  Jensen, S.~Koseki, N.~Matsunaga, K.A. Nguyen, S.~Su, T.L. Windus, M.~Dupuis,
  and J.A. Montgomery.
\newblock General atomic and molecular electronic structure system.
\newblock {\em Journal of Computational Chemistry}, 14:1347--1363, 1993.

\bibitem{bchla-data}
Bacteriochlorophyll a absorption and fluorescence spectra.
\newblock \url{https://omlc.org/spectra/PhotochemCAD/html/135.html}.
\newblock PhotochemCAD, Accessed: 2018-10-09.

\bibitem{sunlight-data}
Standard solar spectra.
\newblock https://pveducation.org/es/pvcdrom/appendices/standard-solar-spectra.
\newblock PVEducation, Accessed: 2018-10-09.

\bibitem{rosencher-vinter}
E.~Rosencher and B.~Vinter.
\newblock {\em Optoelectronics}.
\newblock Cambridge University Press, 2004.

\bibitem{bailey-nie}
{R.E.} Bailey and S.~Nie.
\newblock Alloyed semiconductor quantum dots: Tuning the optical properties
  without changing the particle size.
\newblock {\em J. Am. Chem. Soc.}, 125:7101, 2003.

\bibitem{madelung}
O.~Madelung.
\newblock {\em Semiconductor Data Handbook}.
\newblock Springer, 3rd edition, 2004.

\bibitem{bastard}
G.~Bastard.
\newblock {\em Wave mechanics applied to semiconductor heterostructures}.
\newblock Les {\'E}dition de Physique, 1992.

\bibitem{hinuma-etal}
Y.~Hinuma, A.~Gr{\"u}neis, G.~Kresse, and F.~Oba.
\newblock Band alignment of semiconductors from density-functional theory and
  many-body perturbation theory.
\newblock {\em Physical Review B}, 90:155405, 2014.

\bibitem{wei-etal}
S.~B.~Zhang Su-Huai~Wei and Alex Zunger.
\newblock First-principles calculation of band offsets, optical bowings, and
  defects in cds, cdse, cdte, and their alloys.

\end{thebibliography}
\end{document}